# Heterogeneous reactions between ions $NH_3^+$ and $NH^+$ and hydrocarbons adsorbed on a tungsten surface. Formation of $HCN^+$ in $NH^+$-surface hydrocarbons collisions.

Martina Harnisch[a], Paul Scheier[a], and Zdenek Herman*[a,b]

[a] *Institut for Ion Physics and Applied Physics, University of Innsbruck, Technikerstr. 25, 6020 Innsbruck, Austria*

[b] *J. Heyrovský Institute of Physical Chemistry, v.v.i., Academy of Sciences of the Czech Republic, Dolejškova 3, 18223 Prague 8, Czech Republic*

Abstract

Interaction of $NH_3^+$ ($ND_3^+$) and $NH^+$ with a hydrocarbon-covered tungsten surface kept at room temperature and heated to 150°C and 300°C showed a series of reactions between the projectile ion and hydrocarbons adsorbed on the tungsten surface. Collisions with $NH_3^+$ and particularly with $ND_3^+$ showed formation of $NH_4^+$, $HCNH^+$, and $CH_2NH_2^+$ and $ND_3H^+$, $HCND^+$, and $CH_2ND_2^+$, respectively, with $NH_4^+$ ($ND_3H^+$) strongly prevailing at low incident energies of the projectile ion. No formation of $HCN^+$ ($DCN^+$) could be positively identified. In reactions with $NH^+$ formation of $HCN^+$ was clearly observed; the dependence of the $HCN^+$ normalized ion yield on incident energy seems to indicate a threshold at about 40 eV which may be due to an activation energy or an endothermicity of the surface reaction of about 2.4-3.2 eV.

## 1. Introduction

In our previous paper [1] we reported on formation of $HCN^+$ in heterogeneous reactions between ions $N^+$ and $N_2^+$ with hydrocarbons adsorbed on surfaces of tungsten, beryllium and carbon (carbon-fibre composite). To the best of our knowledge, this was the first experimental observation of a heterogeneous reaction between gas phase ions and surface hydrocarbons which may be of astrophysical interest (formation of HCN in chemical sputtering of carbon films by 150 eV $N_2^+$ was described earlier [2]). The energy of the incident ions varied between 15 and 100 eV and the reactions were investigated at three surface temperatures (room temperature - RT, 150°C, and 300°C). Different surfaces were used to prove that the reaction partner were the hydrocarbons adsorbed on the surface. The reaction with $N^+$ indicated an exothermic reaction with no activation barrier, likely to occur even at very low collision energies. On the other hand, in the reaction with $N_2^+$, the formation of $HCN^+$ was observed only at $N_2^+$ incident energies above 50 eV. This observation, having in mind the earlier established finding of the degree of translation-to-internal energy transfer of 6-8% [3], suggested an activation barrier or reaction endothermicity of about 3-3.5 eV.

In this communication we report on collisions between $NH_3^+$ (and $ND_3^+$) and $NH^+$ with hydrocarbons adsorbed on a surface of tungsten. The measurements were carried out at incident energies of the projectile ions ranging between 15 and 100 eV and with the surface kept at room temperature, 150°C, and at 300°C. Several chemical reactions were observed for both $NH_3^+$ ($ND_3^+$) and $NH^+$ projectile ions, however, $HCN^+$ formation was registered only in collisions between $NH^+$ and surface hydrocarbons.

## 2. Experimental

The experiments were carried out on the tandem apparatus BESTOF in Innsbruck, described in detail in our previous publications (e.g. [4]). It consists of two mass spectrometers arranged in tandem geometry (see Fig. 1). Projectile ions were produced in a Nier-type ion source by electron ionization (energy of 75 - 95 eV) of argon or ammonia ($NH_3^+$ or $ND_3^+$). The ions produced were extracted from the ion source region and accelerated to 3 keV for mass and energy analysis by the double-focusing two-sector-field mass spectrometer. After passing the mass spectrometer exit slit, the ions were refocused by an Einzel lens and decelerated to the required incident energy, before interacting with the surface of the target. The incident impact angle of the projectile beam was kept at 45° and the scattering angle was fixed at 46° (with respect to the surface). The energy spread of the projectile ion beam was about 0.5 eV (full width at half maximum). A fraction of the product ions formed at the surface left the shielded chamber through a 1 mm diameter orifice. The ions were then subjected to a pulsed deflection-and-acceleration field that initiated the time-of-flight analysis of the ions. The second mass analyzer was a linear time-of-flight (TOF) mass spectrometer with an approximately 80 cm long flight tube. The mass-selected ions were detected by a double-stage multi-channel plate connected to a multi-channel scaler and a computer. The product ion yields were obtained by integrating the area under the recorded peaks in the mass spectra.

The pressure in the ion source was 4-8 ×$10^{-5}$ mbar, the bakeable surface chamber and the TOF analyzer were maintained under ultra-high vacuum conditions ($10^{-9}$ mbar) by two turbo-pumps. However, even these ultra-high vacuum conditions did not exclude deposition of a layer of hydrocarbons on the surface, whenever the valve between the sector-field mass spectrometer and the surface chamber was opened and the pressure in the surface region increased to the 1-2×$10^{-8}$ mbar range. In order to investigate the dependence of the product ion signals on the surface temperature, the surface samples could be heated during measurements up to 450°C by a heating wire located inside the surface sample holder.

The hydrocarbons adsorbed on the surface of the solid samples are assumed to be cracked pump oil aliphatic hydrocarbons of chain length of about C7-C8 [5]. The surface of the sample is not homogeneous; it can be viewed as a surface covered largely or partially with islands of hydrocarbons which decrease in size with increasing temperature of the sample. Bombardment of the room-temperature surface by $Ar^+$ carried out after the experiments with ammonia showed traces of ammonia present on the surface.

While in the previous paper [1] we took great care to show that the surface partner in the heterogeneous reactions were indeed surface hydrocarbons by using different metal surfaces, here we limited ourselves only to the tungsten surface. The tungsten sample was a 0.25 mm thick tungsten foil (Alfa Aesar, #10415). Before introduction into the apparatus, the sample was cleaned in three steps, each of 15 min duration: in an ultra-sonic bath with acetone, in boiling water, and again in the ultra-sonic bath with methanol. Subsequent pumping overnight ensured vacuum conditions during measurements in the above-mentioned order.

## 3. Results

### *3.1. Collisions with $NH_3^+$ and $ND_3^+$*

Fig. 2 and 3 show examples of mass spectra of product ions from collisions between $NH_3^+$ ($ND_3^+$) and the room-temperature tungsten sample at incident energies of 30 eV and 50 eV, respectively. Besides the sputtered hydrocarbon ions, known from cross check measurements with $Ar^+$, at m/z 15 ($CH_3^+$), 27 ($C_2H_3^+$), 29 ($C_2H_5^+$), 39($C_3H_3^+$), 41($C_3H_5^+$), 43($C_3H_7^+$) and traces of alkali ions at m/z 23 ($Na^+$) and contributions to m/z 39 and 41 ($K^+$), segregating from the bulk surface material by collision-enhanced thermionic emission. The low ionization energy of the alkali ions leads to unproportionally high ion yield. The spectra with $NH_3^+$ show a strong contribution at m/z 18 ($NH_4^+$), and an increase of ion yields at 28 ($HCNH^+$), and 30 ($CH_2NH_2^+$).

Collisions with $ND_3^+$ enable a closer insight into the product ions of the collisions. The ion yield at m/z 18 ($NH_4^+$) moved entirely to m/z 21 indicating formation of $ND_3H^+$ in a reaction of H-atom transfer from surface hydrocarbons to the projectile ion $ND_3^+$. The ion yield at m/z 30 decreased and the yield at m/z 32 increased strongly suggesting formation of $CH_2ND_2^+$ in a reaction between $ND_3^+$ and surface $CH_3$-hydrocarbon groups. Also, the ion yield at m/z 28 strongly decreased and the ion yield at m/z 29 increased, suggesting formation of $HCND^+$ (presumably deuterated hydrogen cyanide); however, the ion signal at m/z 29 is composed of this possible contribution and the sputtered hydrocarbon ion $C_2H_5^+$. Hence the identification required subtraction of this background as obtained in measurements with $Ar^+$. The ion current at m/z 19 in collisions with $ND_3^+$ is presumably the fragment ion $ND_2H^+$ resulting from dissociation of the abundant $ND_3H^+$. The ion yields at m/z 17 with $NH_3^+$ and 20 ($ND_3^+$) are inelastically scattered projectile ions.

Fig. 4 shows the normalized yields of product ions $ND_3H^+$ (m/z 21), $CH_2ND_2^+$ (m/z 33), and $HCND^+$ (m/z 29, hydrocarbon background subtracted), i.e. the respective yield divided by the sum of all ion yields in the spectrum (yields of alkali ions $Na^+$ and $K^+$ were not taken into consideration), $Y/\Sigma Y$, in dependence on the incident energy of the projectile $ND_3^+$ at two temperatures of the sample (room temperature and 150°C).

The yield of the product ion $NH_4^+$ ($ND_3H^+$) decreased dramatically with increasing incident energy from a yield dominating the spectra at 15-20 eV to a value comparable to the yields of the two other above mentioned product ions. The normalized yield of $HCND^+$ rose from a threshold at 50 eV to a maximum at 70 eV and then decreased to about 20-30% of its peak value at 100 eV. The normalized yield of $CH_2ND_2^+$ increased from a low value at 15-20 eV to a maximum at about 50 eV and then decreased to a very small value at 100 eV.

An insight into the energetics of the reactions can provide a comparison with the respective gas phase reactions of the species involved. Unfortunately, no reactions of $NH_3^+$ with aliphatic hydrocarbons have been reported, with the exception of reactions with methane, acetylene, ethylene, and $C_3H_6$ (no reaction in this case) [6]. However, one can consider the energetics of possible reactions [7]. In the following, n-$C_7H_{16}$ serves as an approximate representative of the hydrocarbons adsorbed on the surface.

The gas-phase reaction with the linear hydrocarbon n-$C_7H_{16}$

$$NH_3^{+\bullet} + C_7H_{16} \rightarrow NH_4^+ + {}^\bullet C_7H_{15} \tag{1}$$

is exothermic by 1.0 eV.

An analogous gas phase reaction leading to $CH_2NH_2^+$

$$NH_3^{+\bullet} + C_7H_{16} \rightarrow CH_2NH_2^+ + H_2 + {}^\bullet C_6H_{13} \tag{2}$$

is endothermic only by 0.35 eV and this small endothermicity, if holding for the surface reaction, may be overcome by a translational-to-internal energy conversion. From earlier measurements [3] on analogous surfaces the translational-to-internal energy conversion in a collision is less than 10%, usually about 6-8% of the incident energy of the projectile. For surface reaction analogous to (2) the incident energy threshold would be at about 5 eV and this threshold is in agreement with the decrease of the $CH_2ND_2^+$ curves in Fig. 4

A possible gas phase reaction leading to $HCNH^+$

$$NH_3^{+\bullet} + C_7H_{16} \rightarrow HCNH^+ + H_2 + {}^\bullet C_6H_{13} \tag{3}$$

is rather strongly endothermic by about 2.4 eV. With the translation-to-internal energy conversion of 6-8% [3], the respective threshold should be then at the incident energy of about 40-30 eV and this is in approximate agreement with the observed threshold of 50 eV for $HCND^+$ in Fig. 3.

A reaction leading to $HCN^+$

$$NH_3^{+\bullet} + C_7H_{16} \rightarrow HCN^+ + 2\,H_2 + {}^\bullet C_6H_{13} \tag{4}$$

is strongly endothermic by about 4.6 eV. A more involved analysis of the ion yields at m/z 27-30 did not show any conclusive indication that $HCN^+$ was formed in $NH_3^+$ ($ND_3^+$) – surface hydrocarbon collisions.

*3.2. Collisions with NH+*

Fig. 5 shows examples of mass spectra from collisions of $NH^+$ (at an incident energy of 70 eV) with the room-temperature surface of tungsten and with the tungsten surface heated to 300°C. At room temperature the spectrum showed only sputtered ions of hydrocarbons at m/z 15 ($CH_3^+$), 27-30 ($C_2H_3^+$ - $C_2H_6^+$), and 39-43 ($C_3H_3^+$ - $C_3H_7^+$) and traces of alkali ions at m/z 23 ($Na^+$) and a contribution to m/z 39 ($K^+$). However, the abundance of $C_2H_3^+$ at m/z 27 was consistently larger than in the spectra with incident $Ar^+$.

At the surface temperature of 300°C, the spectrum showed large amounts of alkalis at m/z 23 ($Na^+$) and m/z 39 and 41 ($K^+$), small amounts of $C_2$ hydrocarbon ions at m/z 27-29, and a strong increase of the abundance of m/z 27, much more than expected for $C_2H_3^+$. Evidently, this results from the formation of $HCN^+$ in $NH^+$ - surface hydrocarbon collisions. The standard subtraction (as described in [1]) of the hydrocarbon background at m/z 27, $Y(C_2H_3^+)$, obtained by comparing the $NH^+$ spectra with those from $Ar^+$ collisions, led to the net $HCN^+$ signal at this m/z. Fig. 6 shows the data presented as normalized yields, $\Delta Y = Y(27) - Y(C_2H_3^+)$ divided by the sum of all ion yields in the spectrum (excluding alkali ions $Na^+$ and $K^+$), $\Delta Y/\Sigma Y$.

No gas phase reactions of $NH^+$ with aliphatic hydrocarbons have been reported, with the exception of reactions with methane and ethylene [6].

A possible reaction

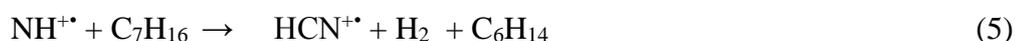
$$NH^{+\bullet} + C_7H_{16} \rightarrow HCN^{+\bullet} + H_2 + C_6H_{14} \qquad (5)$$

is exothermic by about 3 eV and thus a surface reaction between $NH^+$ and a terminal $CH_3$ group of the surface-adsorbed hydrocarbons should be likely to occur. However, the curves of normalized yields of $HCN^+$ in Fig. 5 indicate an increase only at incident energies of $NH^+$ above about 40 eV. This finding would imply an activation barrier or an endothermicity of the surface reaction leading to $HCN^+$ of about 2.4-3.2 eV. One can only speculate of the reason of this behavior. First of all, reaction (5) may not be representative of the surface reaction and the neutral products may be different (e.g., "healing" of the defect in the hydrocarbon terminal group, and forming a shortened saturated hydrocarbon chain may not take place). Indeed, another possible reaction

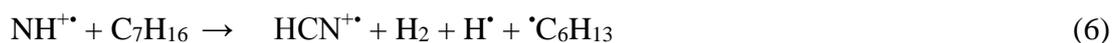
$$NH^{+\bullet} + C_7H_{16} \rightarrow HCN^{+\bullet} + H_2 + H^{\bullet} + {}^{\bullet}C_6H_{13} \qquad (6)$$

is endothermic by 1.3 eV, closer to the value derived for the threshold of the surface reaction. Second, the reaction requires considerable rearrangement of the constituents in the $NH^+$-surface hydrocarbon interaction and this may be connected with some activation energy.

**4. Conclusions**

1. Interaction of $NH_3^+$ ($ND_3^+$) and $NH^+$ with a tungsten surface covered with hydrocarbons and kept at room temperature or heated to 150°C and 300°C showed a series of reactions between the projectile ion and hydrocarbons adsorbed on the tungsten surface.
2. With the projectile ion $NH_3^+$ surface reactions leading to $NH_4^+$, $HCNH^+$ and $CH_2NH_2^+$ were observed, studies with $ND_3^+$ enabled determining wherefrom the hydrogen/deuterium in the product ions came: the product ions were $ND_3H^+$ (H-atom transfer from surface hydrocarbons), $HCND^+$ (CH from surface hydrocarbons), and $CH_2ND_2^+$ ($CH_2$ from surface hydrocarbons).
3. The dependence of normalized ion yields on the incident energy of the projectiles suggests a strong preference to form $ND_3H^+$ at low incident energies, decreasing strongly between 30-60 eV, a threshold of 50 eV for $HCND^+$ production (corresponding to endothermicity or an activation barrier of about 3-4 eV, approximately consistent with an estimation of the endothermicity of an analogous gas phase reaction of 2.4 eV), and a very low threshold for formation of $CH_2ND_2^+$ below 10 eV (consistent with an estimation of the possible surface reaction of about 0.35 eV).
4. No reaction leading to $HCN^+$ in $NH_3^+$-surface hydrocarbons interaction could be positively identified.
5. In reactions between $NH^+$ and surface hydrocarbons formation of $HCN^+$ was clearly identified. The dependence of the normalized $HCN^+$ ion yield on incident energy seems to indicate a threshold at about 40 eV which may be due to an activation energy or an endothermicity of the reaction of about 2.4-3.2 eV.

**Acknowledgments**

This work, supported by the European Communities under the Contracts of Association between EURATOM ÖAW and EURATOM IPP.CR, was carried out within the framework of the European Fusion Development Agreement (EFDA). The views and opinions expressed herein do not necessarily reflect those of the European Commission. The research was partly supported by FWF, Wien, Project P26635.

FIGURE CAPTIONS

**Fig. 1:** Schematics of the tandem apparatus BESTOF: A mass-selected ion beam interacts with the surface, product ions are detected by a time-of-flight (TOF) mass spectrometer.

**Fig. 2:** Mass spectrum of product ions from the interaction of $NH_3^+$ (red) and $ND_3^+$ (black) with the hydrocarbon-covered tungsten surface kept at room temperature; incident energy of the projectile ions was 30 eV.

**Fig. 3:** Mass spectrum of product ions from the interaction of $NH_3^+$ (red) and $ND_3^+$ (black) with the hydrocarbon-covered tungsten surface kept at room temperature; incident energy of the projectile ions was 50 eV.

**Fig. 4:** Dependence on the incident energy of the normalized ion yields, $Y/\Sigma Y$, of the product ions $ND_3H^+$, $HCND^+$, and $CH_2ND_2^+$ from collisions of $ND_3^+$ with surface hydrocarbons adsorbed on a tungsten surface kept at room temperature RT (a) and at 150°C (b); left-hand scale refers to normalized ion yield of $ND_3H^+$, right-hand scale to $HCND^+$, and $CH_2ND_2^+$ (see arrows).

**Fig. 5:** Examples of mass spectra of product ions from interactions of $NH^+$ with surface hydrocarbons adsorbed on the tungsten surface kept at room temperature and 300°C. Incident energy of the projectile ions $NH^+$ was 70 eV.

**Fig. 6:** Dependence on the incident-energy of the normalized net ion yields, $\Delta Y/\Sigma Y$, of the product ion $HCN^+$ from collisions of $NH^+$ with surface hydrocarbons adsorbed on a tungsten surface kept at room temperature (RT, top) and at 300°C (bottom); the dashed line $Y(27)/\Sigma Y$ shows the total normalized yield of m/z 27 before correction for the $C_2H_3^+$ background.

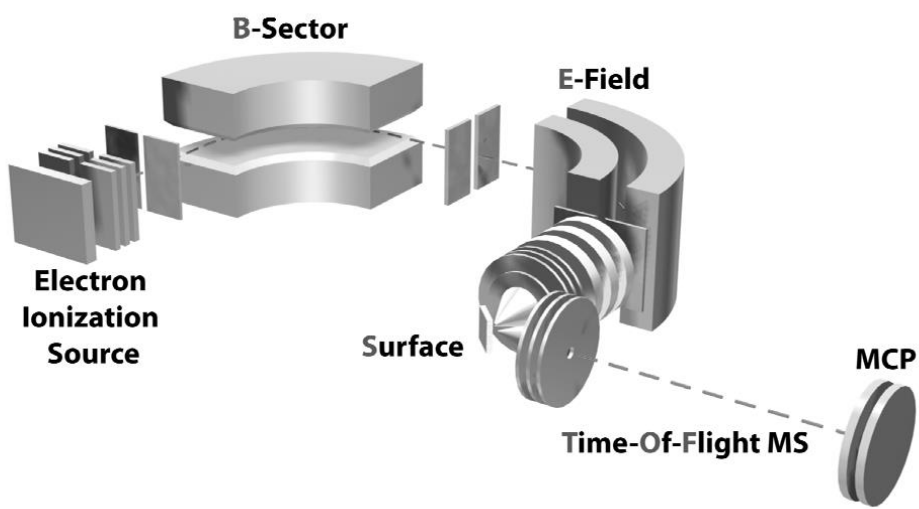

**Fig. 1**

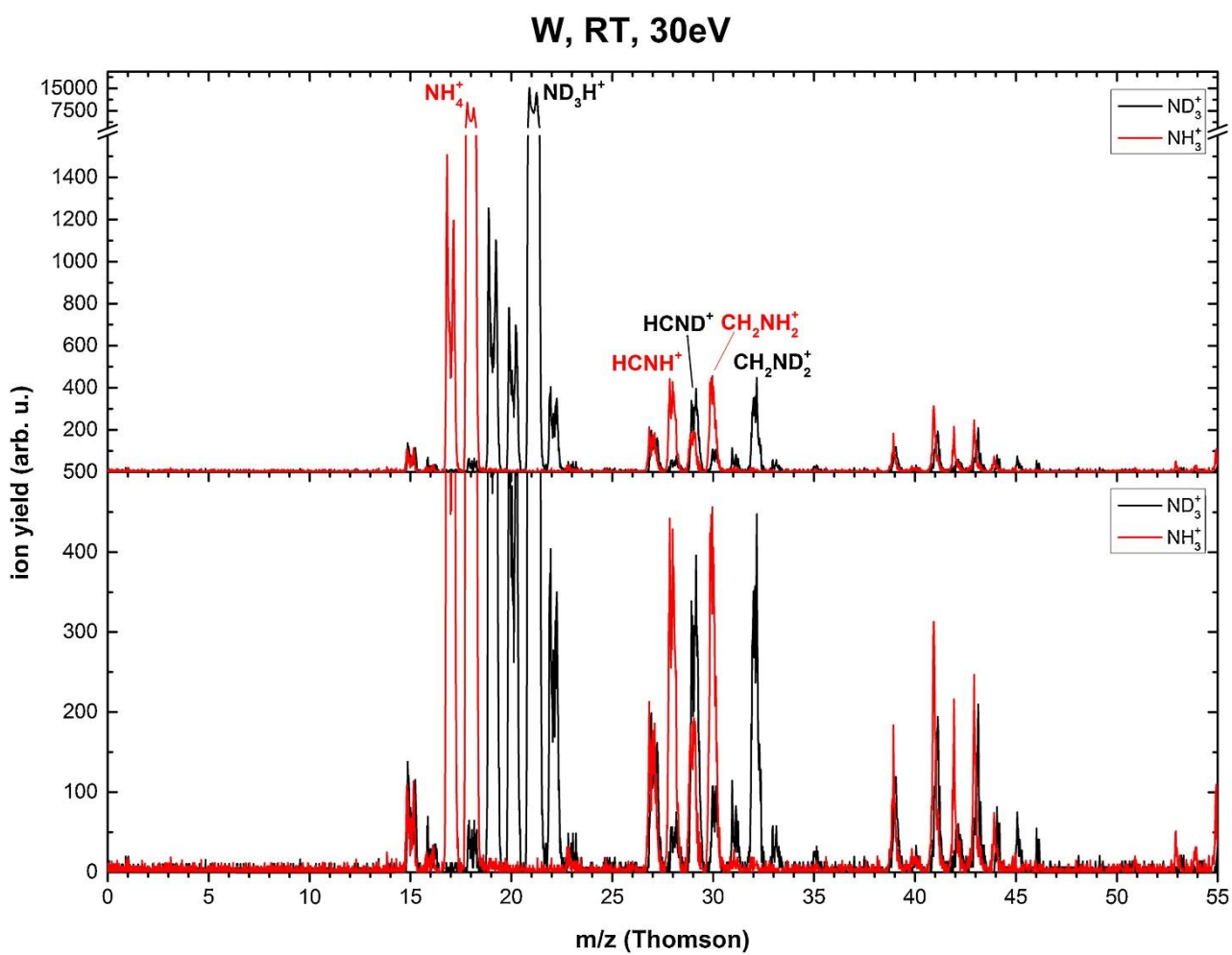

**Fig. 2**

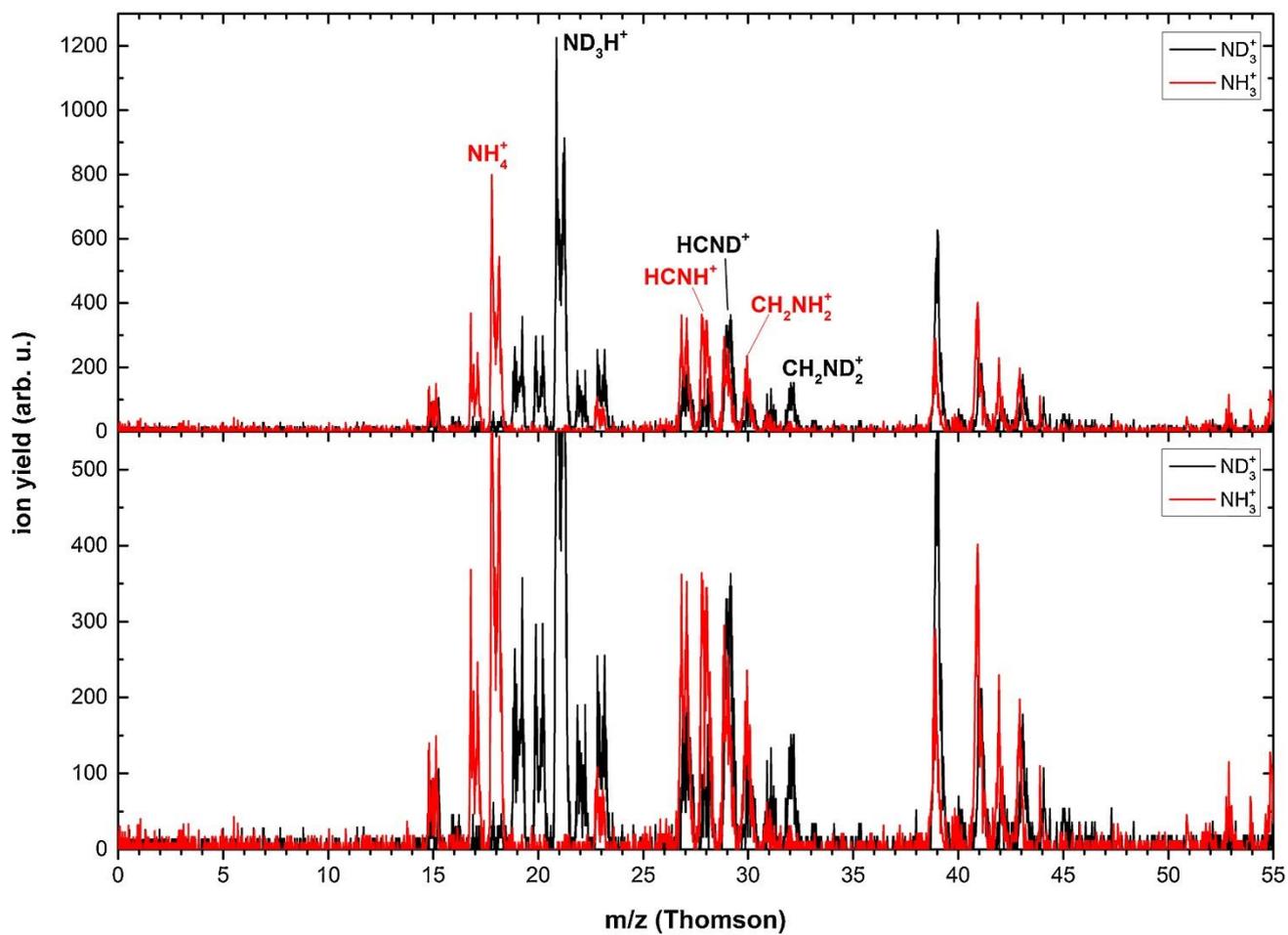

**Fig. 3**

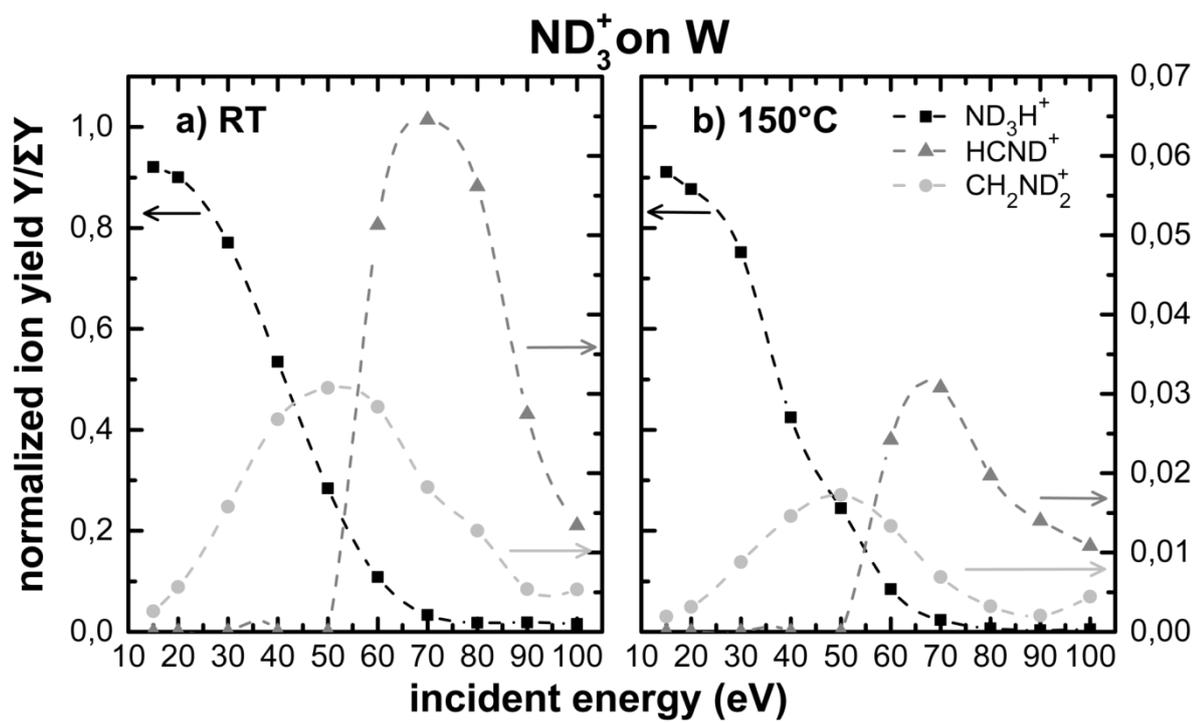

**Fig. 4**

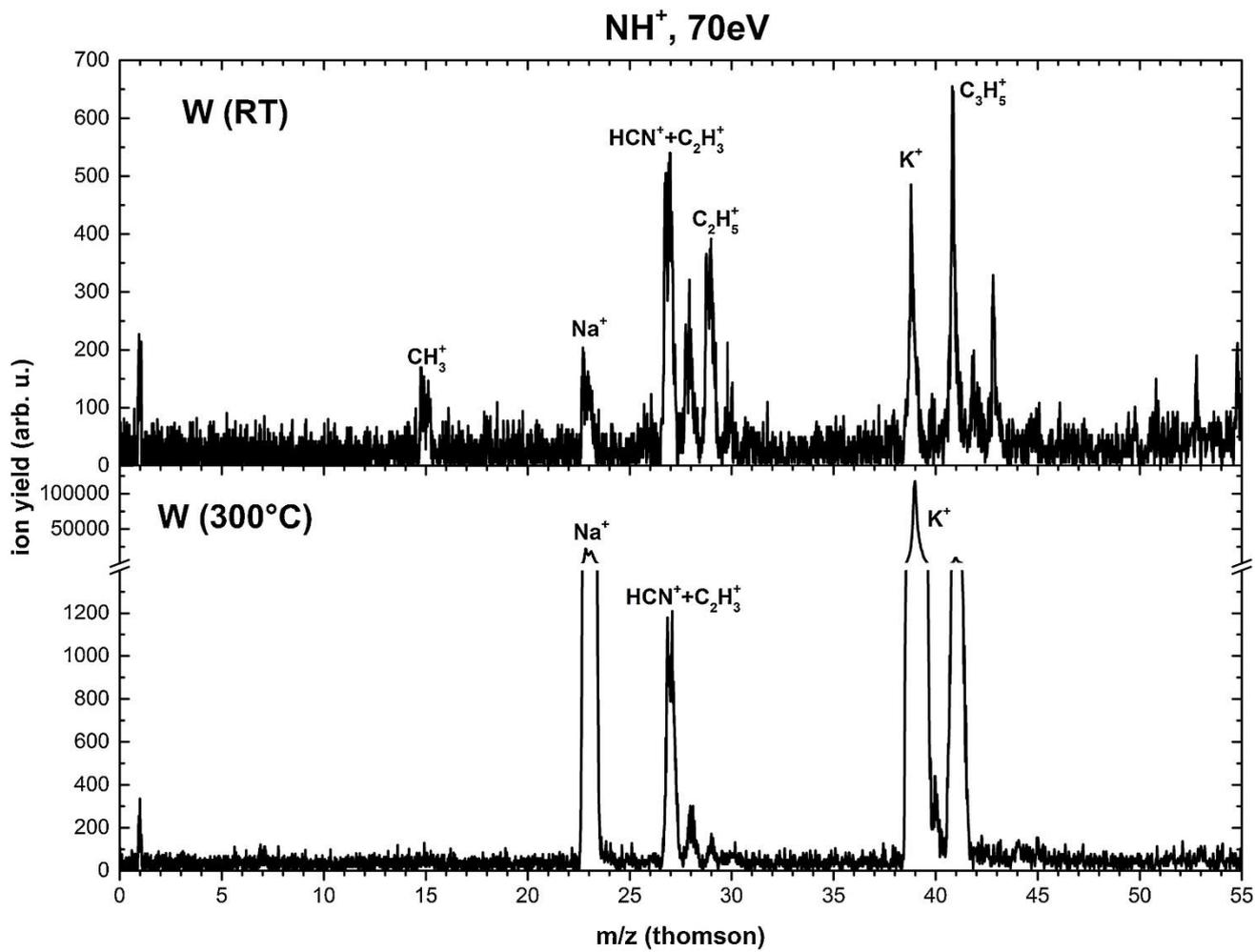

**Fig. 5**

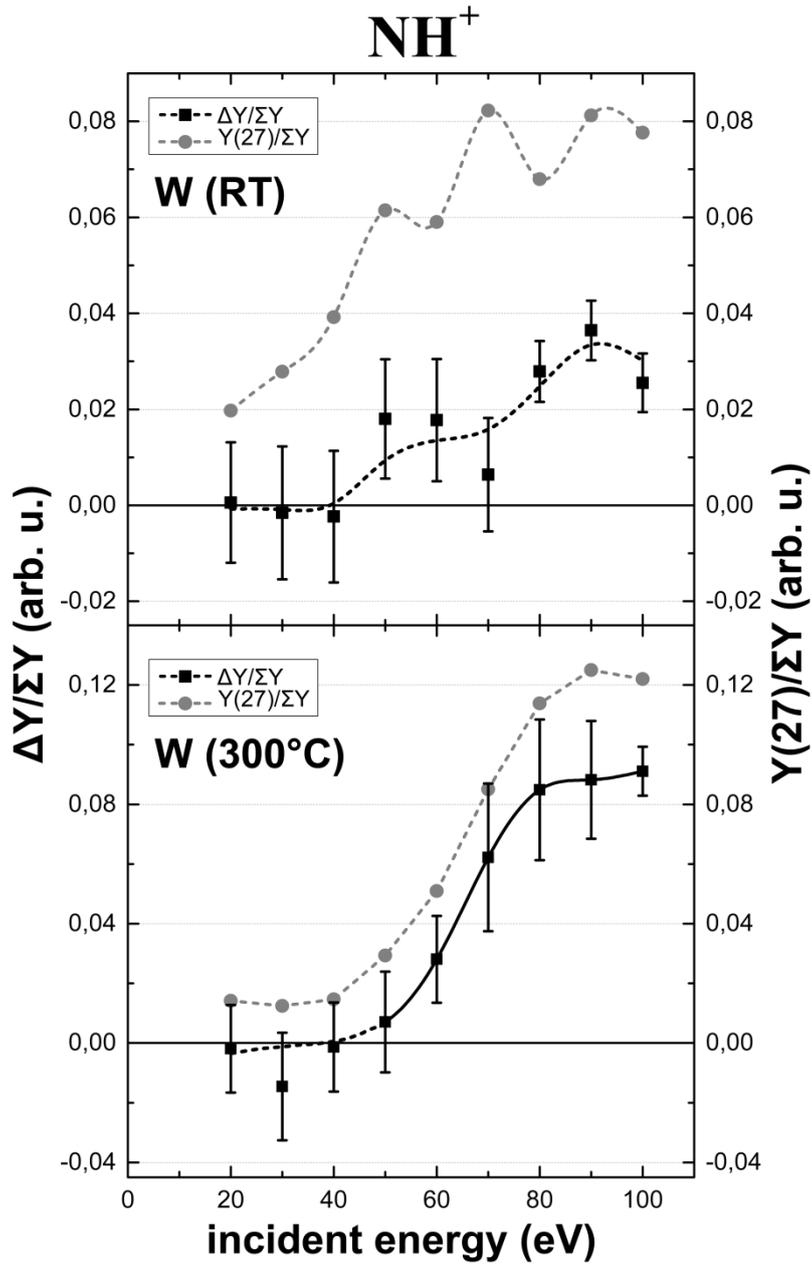

**Fig. 6**